\documentclass[aps,pre,reprint,superscriptaddress,showpacs]{revtex4-1}

\usepackage[pdftex]{graphicx}
\setkeys{Gin}{clip=true,draft=false}
\DeclareGraphicsExtensions{.pdf}
\usepackage[cmex10]{amsmath}
\usepackage{amsthm}
\usepackage{amsfonts}
\usepackage[cspex,bbgreekl]{mathbbol}
\newcommand{\R}{{\ifmmode\mathbb{R}\else$\mathbb{R}$\fi}}
\newcommand{\K}{{\ifmmode\mathbb{K}\else$\mathbb{K}$\fi}}
\newcommand{\N}{{\ifmmode\mathbb{N}\else$\mathbb{N}$\fi}}
\newcommand{\C}{{\ifmmode\mathbb{C}\else$\mathbb{C}$\fi}}

\newcommand{\fracd}[2]{\frac{\displaystyle #1}{\displaystyle #2}}
\newcommand{\partialder}[2]{\frac{\displaystyle\partial #1}{\displaystyle\partial #2}}

\newcommand{\partialdersec}[2]{\frac{\displaystyle\partial^2 #1}{\displaystyle\partial #2^2}}

\newcommand{\norm}[1]{{\left\| #1 \right\|}}

\newcommand{\ket}[1]{\ensuremath{|#1\rangle}} 
\newcommand{\bra}[1]{\ensuremath{\langle#1|}}
\newcommand{\braket}[2]{\langle#1|#2\rangle}
\newcommand{\brainket}[3]{\langle#1|#2|#3\rangle}

\usepackage{color}
 
\begin{document}

%\preprint{}

% \title{Analytical injection, absorption and remapping for efficient solutions of time dependent Schr\"{o}dinger equation}
\title{Absorption and Injection Models for Open Time-Dependent Quantum Systems}


\author{F. L. Traversa}
\email[]{fabio.traversa@polito.it}
%\thanks{}
\affiliation{Departament d'Enginyeria Electr\`onica, Universitat Aut\`onoma de Barcelona, 08193-Bellaterra (Barcelona), Spain}

\author{Z. Zhan}
\email[]{zhenzhanh@gmail.com}
%\thanks{}
\affiliation{Departament d'Enginyeria Electr\`onica, Universitat Aut\`onoma de Barcelona, 08193-Bellaterra (Barcelona), Spain}

\author{X. Oriols}
\email[]{xavier.oriols@uab.es}
%\thanks{}
\affiliation{Departament d'Enginyeria Electr\`onica, Universitat Aut\`onoma de Barcelona, 08193-Bellaterra (Barcelona), Spain}


\date{\today}

\begin{abstract}
In the time-dependent simulation of pure states dealing with transport in open quantum systems, the initial state is located outside of the active region of interest. Using the superposition principle and the analytical knowledge of the free time-evolution of such state outside the active region, together with absorbing layers and remapping, a model for a very significant reduction of the computational burden associated to the numerical simulation of open time-dependent quantum systems is presented. The model is specially suited to study (many-particle and high-frequency effects) quantum transport, but it can also be applied to any other research field where the initial time-dependent pure state is  located outside of the active region. From numerical simulations of open quantum systems described by the (effective mass) Schr\"{o}dinger and (atomistic) tight-binding equations, a reduction of the computational burden of about two orders of magnitude for each spatial dimension of the domain with a negligible error is presented.
\end{abstract}

% General formulation of transport theory}
%\keywords{}

\maketitle

\section{Introduction}

%%%%%%%%%%%%%%%%%%%%%%%%%
%%%%%%%%%%%%%%%%%%%%%%%%%
%%%%%%%%%%%%%%%%%%%%%%%%%
%%%%%%%%%%%%%%%%%%%%%%%%%

The ultimate reason why the quantum theory gives rise to a host of puzzling and fascinating phenomena (without classical counterpart) is because quantum states live in a high-dimensional and abstract configuration space (rather than in the ordinary 3D physical space). The computational burden associated with the $N$-particle state makes the exact solution of the many-particle Schr\"{o}dinger equation inaccessible in most practical situations. Historically, among other strategies, the computational burden has been reduced by selecting Hamiltonian eigenstate as the representation of particles. For example, the (lowest energy) ground state successfully explains the behavior of equilibrium quantum systems.

However, there are many quantum scenarios where the time-dependent Schr\"{o}dinger equation needs to be explicitly considered \cite{Robinett_2011}. For example, when light intensity is sufficiently small, a first-order perturbative theory is enough to describe the main features of the interaction between light and matter, but when the light intensity becomes larger, a plethora of different phenomena appears and more accurate models are required. The exact quantum description of the photoionisation due to the interaction of an atom (or molecule) with a (classical) electromagnetic pulse in the non-relativistic regime is the time-dependent Schr\"{o}dinger equation \cite{Domokos_2002,ruiz_lithium_2005,krammer_2006,picon_transferring_2010,wu_bohmian-trajectory_2013}. Equivalently, the quantum transport in mesoscopic systems has been mainly understood from (time-independent) scattering states  \cite{Buttiker2,Buttiker3}. However, strictly speaking, the scattering states do not belong to the \textit{physical} states of any Hilbert space because they cannot be normalized to unity. In other words, strictly speaking, these states cannot be associated to an electron localized at the right or left of the device active region because they extend everywhere, at any time \cite{footnote3}. Certainly, these Hamiltonian eigenstates can be used as a base to define well-localized electrons by superposition. However, a proper superposition of eigenstates can only be useful numerically to describe the evolution of wave packets in time-independent Hamiltonians (where eigenstates remain invariant with time). Any time-dependent potential requires an explicit solution of the time-dependent Schr\"{o}dinger equation. 

The need for time-dependent algorithms to properly understand quantum transport has already been discussed in the literature in several different contexts. For example, the time-independent density functional theory is said to be unable to properly capture non-equilibrium scenarios, while time-dependent versions are mandatory for successful predictions \cite{rungePRL84,TDDFTcurrent,Angel,errorDFT}. Similarly, in quantum transport, it is said that the Landauer formula is incomplete because one-particle scattering probabilities do not capture the many-body effects \cite{diVentra}. In the same way, time-independent pictures has many difficulties to treat AC and transients dynamics properly \cite{Niehaus,Guillem,Ferry}. Additionally, the advantages of modeling transport in waveguides using wave packets have also been indicated \cite{Robinett_2011,Domokos_2002,Das_2011,krammer_2010}.  We have also shown quite recently that the Bohmian conditional wave function is a very powerful tool to deal with both quantum many-body problems and non-unitary evolutions \cite{Guillem, Oriols} and useful to simulate AC and transient current as well as noise in mesoscopic devices \cite{11_Traversa,Oriols_book}. By constructions, such (Bohmian conditional) wave functions do also require a time-dependent evolution. 

\subsection{Problem setting}

\begin{figure}
\centerline{
\includegraphics[width=\columnwidth]{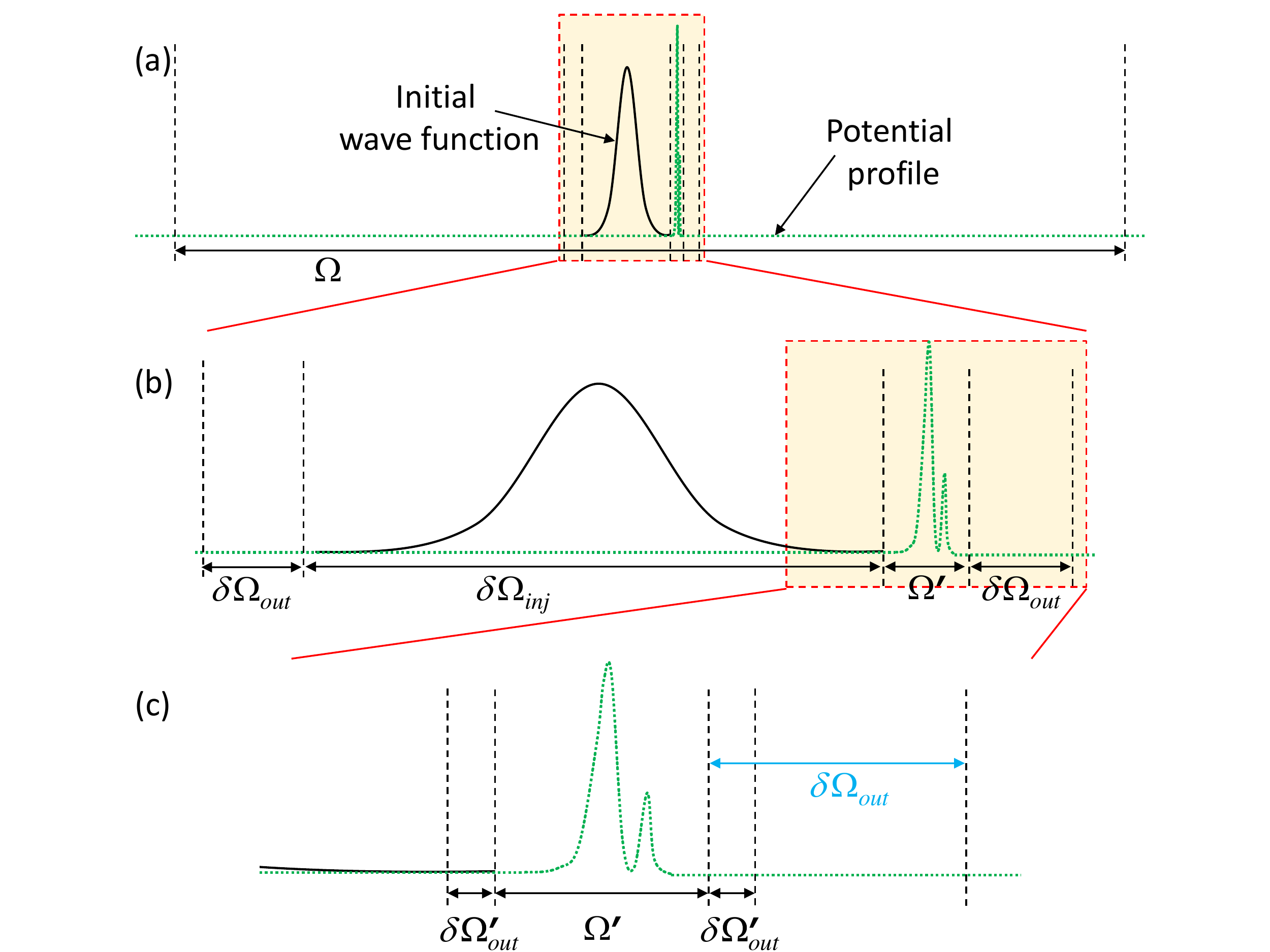}}
\caption{\label{domain} (Color online) Spatial simulation domains. (a) \textit{``Infinite'' domain}: full domain $\Omega$ is a large enough domain to avoid interactions with boundaries during the simulation time. (b) \textit{Absorbing layers}: reduced domain $\delta\Omega_{out}\cup\delta\Omega_{inj}\cup \Omega\rq{}\cup \delta\Omega_{out}$ using standard absorbing algorithms where $\delta\Omega_{inj}$ is the layer for the injection, i.e., the layer that includes the wave function at the initial time, $\Omega\rq{}$ is the interaction box and $\delta\Omega_{out}$ at both sides are the absorbing layers. (c) \textit{Absorbing layers plus analytical injection}: further reduced domain $\delta\Omega_{out}\rq{}\cup \Omega\rq{}\cup \delta\Omega_{out}\rq{}$ presented in this work; it has no need of injection layer and employs smaller absorbing layers $\delta\Omega_{out}\rq{}\ll \delta\Omega_{out}$ obtained exploiting a change of coordinates (remapping).}
\end{figure}

The main motivation of the present work is reducing the computational burden associated to the study of quantum transport with time-dependent pure states. As we will see, the computation of quantum transport has some peculiarities that imply new and unexplored methods to greatly simplify the numerical computational resources.  A general scenario for modeling quantum transport assumes a finite domain $\Omega$ where the time-dependent wave functions is solved. See figure~\ref{domain}--a. Such domain contains a flat potential region except in the interaction box $\Omega\rq{}$, i.e., the so-called active region, where the potential $V$ can be time-dependent and inhomogeneous. By construction, the support of the time-dependent wave function, at large times, can be very far from the interaction box $\Omega\rq{}$ therefore, to eliminate spurious events at the boundaries, $\Omega$ is generally selected extremely  large. Thus, in order to avoid the very large domain of figure~\ref{domain}--a, several absorbing boundary conditions have been developed for the time-dependent Schr\"{o}dinger equation (see \cite{review,04_Muga} and references therein). Some approaches are based, for example, either on fitting the wave function to plane waves at the boundaries \cite{Hadley,Yevick2}, or on time convolution integrals at the boundaries to construct transparent boundary condition \cite{89_Baskakov, 03_arnold, Lubich_02, Jiang_08}. If this approximate solution actually coincides on $\Omega\rq{}$ with the exact solution of the whole-space problem, one refers to these boundary conditions as transparent boundary conditions \cite{review}. However, such transparent boundary conditions require an increment of the  complexity in the computer implementation due to their formulation employing spatial and time-convolution integrals. Other much simpler strategies that provide a negligible error when compared to the transparent boundary conditions are greatly preferred. One common strategy of this second type is the use of absorption or attenuation layers $\delta \Omega_{out}$ at the boundaries of the simulation domain \cite{Kosloff, Yevick, 04_Muga,McCurdy_02} as depicted in figure~\ref{domain}--b. The value of the wave function at $\delta \Omega_{out}$ is decreased, at each time step of the simulation. This idea can be also interpreted as an application of the exterior complex scaling \cite{McCurdy_02} as well as adding an artificial complex potential at $\delta \Omega_{out}$ \cite{Kosloff,04_Muga}. See appendix \ref{appendix1}. Let us notice, however, that this algorithm do still require a quite large domain (see $\delta \Omega_{inj}$ in figure~\ref{domain}--b) to properly define the initial state. 

Many of the above strategies available in the literature have been developed for the 1D case with no easy implementation to higher dimensionality (some expections for 2D extensions involving time convolution integrals with a near optimal complexity can be found in Refs. \cite{Lubich_02, Jiang_08}). In order to apply absorbing boundary conditions in realistic electronic device simulator \cite{Guillem, Oriols,11_Traversa}, one is interested in a algorithm (i) with a negligible increment of the computational effort, (ii) easily generalizable to quantum systems of any dimensionality (1D, 2D and 3D) and (iii) not restricted to the continuous Schr\"{o}dinger equation but applicable also to atomistic tight binding equations (which are nowadays quite common in quantum transport where transport and band-structure phenomena are fully mixed). Among the above methods, the one based on the attenuation layer \cite{Kosloff,04_Muga} schematically represented in figure~\ref{domain}--b fulfills these requirements. However, to the best of our knowledge, in the attenuation layer method (in in fact in all previous works on absorbing boundary conditions for the time-dependent wave function \cite{Kosloff, 04_Muga, Yevick, Hadley, Yevick2, review, 89_Baskakov, 03_arnold, Lubich_02, Jiang_08, McCurdy_02}), the simulation domain is selected so that the support of the initial state perfectly fits inside the domain, i.e., there is an injection layer large enough to contain the whole initial state when applied to transport. See $\delta \Omega_{inj}$ in figure~\ref{domain}--b. Although this condition seems reasonable, we will see in this work that it implies an important computational drawback for time-dependent quantum transport. Indeed, a general scenario for modeling quantum transport assumes a time-dependent inhomogeneous potential in the active region $\Omega\rq{}$ and an homogeneous potential outside. The initial wave function is located outside $\Omega\rq{}$ in $\delta\Omega_{inj}$. See figure~\ref{domain}--b. For example, a typical scenario is a (tunneling) barrier of few nanometres plus an initial wave function located far from the barrier (i.e. outside $\Omega\rq{}$) and whose spatial dispersion is tenths of nanometres (even much larger than the active region itself). The evolution of the initial wave function before impinging with the barrier is quite trivial. Under these circumstances, we demonstrate in section~\ref{section_Injection} that it is possible to avoid the injection layer $\delta \Omega_{inj}$ and reduce the domain to $\delta\Omega_{out}\rq{}\cup\Omega\rq{}\cup\delta\Omega_{out}\rq{}$ as depicted in figure~\ref{domain}--c using a simple and general injection algorithm. Moreover, in section~\ref{section_dumping} we also present a new variant to the absorbing boundaries employing attenuation layers similar to \cite{Kosloff} but exploiting a change of coordinates (we call remapping, see section~\ref{section_remapping}) of the attenuation layer that allows a sensible reduction of the width of attenuation layer itself ($\delta\Omega_{out}\rq{}\ll \delta\Omega_{out}$, figure~\ref{domain}--c) as proved in section~\ref{section_implementation}. These new simulation schemes imply an unprecedented reduction of the computational burden associated to numerical simulations of time-dependent wave packets. Finally, even if in this work we present the 1D case only for sake of compactness and clarity, the generalization to 2D and 3D dimensions is possible even if not completely trivial due to some issues arising in higher dimensions not fully treated here (errors depend in a complicated way on the angle of incidence). However this work represent the seed for future generalization to higher dimension. 

%%%%%%%%%%%%%%%%%%%%%%%%%%%
%%%%%%%%%%%%%%%%%%%%%%%%%%%
%%%%%%%%%%%%%%%%%%%%%%%%%%%
\bigskip 

\section{General consideration}

%%%%%%%%%%%%%%%%%%%%%%%%%%%
%%%%%%%%%%%%%%%%%%%%%%%%%%%
%%%%%%%%%%%%%%%%%%%%%%%%%%%

We study the time dependent transport of particles (electrons) in a tunneling region.  For the sake of simplicity we consider the 1D system. Below, we present a brief summary of the formalization and of the results of the effective mass and tight binding formulation of the Schr\"{o}dinger equation relevant for this work.

\subsection{The Hamiltonian}\label{section_hamiltonian}
We consider the time-dependent Schr\"{o}dinger equation: 
\begin{equation}
i\hbar\partialder{ \ket {\psi(t)}}{t}=\left(\hat H_0 +\hat U\right) \ket{\psi(t)},
\label{shrod}
\end{equation}
where $\ket {\psi(t)}$ is a state and $\hat H=\hat H_0+\hat U$ is the Hamiltonian in some particular Hilbert space split into the free particle Hamiltonian $\hat H_0$ (i.e. no interactions are included) and the potential operator $\hat U$ representing interaction with external force fields. Given the position state $\ket x$, we can define the wave function $\psi(x,t)=\braket {x}{\psi(t)}$ and its (effective mass) Hamiltonian: 
\begin{equation}
\brainket {x} {\hat H_{em}} {\psi(t)}=-\fracd{\hbar^2}{2m^*}\partialdersec{\psi(x,t)}{x}+U(x,t)\psi(x,t),
\label{h_x}
\end{equation}
where $m^*$ the particle (effective) mass and $U(x,t)$ the external potential. 
Second, since the work is motivated for electronic transport (in crystal materials), we discuss also a particle in the Hilbert space defined by the (1D regularly distributed) M atom positions, $x_j=j\; \Delta x$. The state of the system is defined now as $\psi_j(t)=\braket {j} {\psi(t)}$ and the (1D nearest-neighbor tight-binding) Hamiltonian: 
\begin{equation}
\hat H_{tb}=\sum_{j=1}^{M} \rho \ket {j} \bra {j} + u \ket {j} \bra{j+1} + u \ket {j} \bra{j-1}+U_j(t)\ket {j} \bra {j}  
\label{h_tb}
\end{equation}
where $U_j(t)=U(x_j,t)$ and $\ket {j}$ are the (Wannier) states associated to the $j$-atom. We assume that all $\ket {j}$ form a complete $\sum_{j=1}^M \ket j \bra j= \mathbf{1}$ and orthonormal $\braket i j =\delta_{ij}$ set. It is very enlightening to rewrite \eqref{h_tb} in the $j$-site representation:
\begin{equation}
\brainket {j} {\hat H_{tb}} {\psi}=  u (\psi_{j-1} + \psi_{j+1}) +\rho \psi_{j} + U_j\psi_{j}  
\label{h_tbx}
\end{equation}
where (for compactness) we have not written the time dependence of the state. 
The generalization of \eqref{h_x} and \eqref{h_tbx} to 2D and 3D cases is straightforward and it will be briefly discussed in the conclusions.

\subsection{Hamiltonian Eigenstates and eigenvalues}

Let us consider the free particle Hamiltonian $\hat H_0$. Then, in the effective mass scenario, the Hamiltonian eigenfunctions are plane waves $\ket k_{em}=1/\sqrt{2 \pi}\int \exp(ikx)\ket x dx$ with eignevalues:
\begin{equation}
E_{em}(k)=\frac{\hbar^2 k^2} {2 m^*}
\label{dispersion_em}
\end{equation}
for any value of the wave vector $k$.  

On the other hand, the tight binding $\hat H_0$ has eigenkets $\ket k_{tb} $ of the form of Bloch eigenfunctions: 
\begin{equation}
\ket k_{tb} =  \sum_{j=1}^M e^{i k j \Delta x} \ket j
\label{bloch}
\end{equation}
for $k \in [-\pi/\Delta x,\pi/\Delta x]$ with eigenvalues: 
%\begin{equation}
%\hat H_{tb_0} \ket k = E_{tb}(k) \ket k= (\rho+u\;e^{-ik\Delta x}+u\;e^{ik\Delta x}) \ket k
% E_{tb}(k) = (\rho+u\;e^{-ik\Delta x}+u\;e^{ik\Delta x}) 
%\label{eigen}
%\end{equation}
%These tight-binding eigenvalues $E_{tb}(k)$ are in general:
\begin{equation}
 E_{tb}(k)=\rho + 2\;u\; \cos(k \Delta x).
 \label{dispersion_tb}
 \end{equation}
which represent the so called (energy-wavevector) dispersion relationship. 

\subsection{Localized initial state in a flat potential region}
\label{initialstate}

As mentioned in figure \ref{domain}, the entire quantum domain is \textit{artificially} divided into two reservoirs (left and right) and an interaction box $\Omega\rq{}$. At the initial time,  the wave function of the particle (electron) is fully localized in one of the reservoirs while, at a final time, its probability presence is delocalized into the left or right reservoirs (but not in $\Omega\rq{}$).  The initial state can be written as a proper superposition of Hamiltonian eigenstate, whose time-evolution (inside the reservoir) can be written, in general \cite{Cohen}, as: 
\begin{equation}
\ket {\psi(t)} = \int_{-\infty}^{\infty} a(k) e^{-i E(k) (t-t_0)/\hbar} \ket k dk
\label{superposition_x}
\end{equation}
with $a(k)=\braket {k}{\psi(t_0)}$. For \eqref{h_x}, a very reasonable assumption for computing \eqref{superposition_x} analytically in a flat-potential reservoir is the following Gaussian wave packet:
\begin{align}
\psi_G(x,t)=\left[\fracd{\sigma_{0}^2}{2\pi(\sigma_{0}^4+\sigma^4_x(t))}\right]^{\frac{1}{4}}e^{i[\varphi(t)+k_x(x-x_0)]} \hspace{0.4cm} \nonumber\\\
\times\exp\left[ -\fracd{(x-x_0-2k_x\sigma^2_x(t))^2}{4(\sigma^2_{0}+i\sigma^2_x(t))} \right],
\label{gaussian}
\end{align}
where $\sigma_0$ is the spatial variance of the wave packet at $t=0$, $x_0$ the initial central position, $\sigma^2_x(t)=\hbar\;t/2m^*$, $k_x$ the central wave vector and $\varphi(t)=-\theta(t)- k_x^2\hbar t/2m^*$ with $\theta(t)$ solution of $\sigma_0^2\tan(2\theta)=\hbar t/2m^*$. Moreover, it can be simply verified that $\int_{-\infty}^{\infty}|\psi_G(x,t)|^2dx=1$

Equivalently, the same gaussian wave function can be used as the initial state for the tight-binding model with:  
\begin{align}
\psi_{G_i}(0)=\braket {i} {\psi_G(0)}=\psi_G(x_i,0)/N,
\label{gaussian_tb}
\end{align}
being $N$ a constant for a proper normalization. Strictly speaking, $\psi_{Gi}(0)$ is not a spatial wave function, but a spatial envelope wave function.

%%%%%%%%%%%%%%%%%%%%%%%%%
%%%%%%%%%%%%%%%%%%%%%%%%%
%%%%%%%%%%%%%%%%%%%%%%%%%

\section{Metamathematical algorithms}\label{Metamathematical algorithms}

After interacting in $\Omega'$, the wave-function freely spreed out in the domain $\Omega$ of figure \ref{domain}--a. Our novel model to shorten the simulation box is based on analytical injection, plus absorbing and remapping algorithms. We will see that the simultaneous use of both these two techniques together with the analytical injection provides the shortest simulation box, with a negligible error and a very small additional computational effort. 

%%%%%%%%%%%%%%%%%%%%%%%%%
%%%%%%%%%%%%%%%%%%%%%%%%%
%%%%%%%%%%%%%%%%%%%%%%%%%

\subsection{Analytical injection} \label{section_Injection}

%%%%%%%%%%%%%%%%%%%%%%%%%
%%%%%%%%%%%%%%%%%%%%%%%%%
%%%%%%%%%%%%%%%%%%%%%%%%%

Since we are interested in time-dependent wave-packets whose initial states are localized in the left (or right) reservoir, it would seem that one had to include the layer $\delta \Omega_{inj}$ in figure \ref{domain}--b as an avoidable part of the simulation box. As we discuss in Sec. \ref{initialstate}, the time evolution of a wave function in the $\delta \Omega_{inj}$ is quite predictable, even analytical for some initial states, as for example Gaussian wave packets  (see Eq. \eqref{gaussian}). Therefore, one can envision an algorithm to avoid the explicit consideration of the reservoirs in the simulation box (during the injection process). 
In order to pursue this goal, we present an injection algorithm that can work for both effective mass and tight binding Hamiltonians discussed in this work. 

\subsubsection{State Split}\label{section_state_split}

The state $\ket{\psi}$ solution of eq.~\eqref{shrod} in the whole domain $\Omega$ can be decomposed as
\begin{equation}
\ket{\psi(t)}=\ket{\psi_0(t)}+\ket{\phi(t)},
\label{sum}
\end{equation}
where $\ket{\psi_0}$ is the free particle solution (i.e. the solution of eq.~\eqref{shrod} with $\hat U=0$). Using linearity of Schr\"{o}dinger equation, it can be found that $\ket {\phi (t)}$ is solution of 
\begin{equation}
i\hbar\partialder{\ket{\phi(t)}}{t}=\left(\hat H_0  +\hat U\right)\ket{\phi(t)}+\hat U \ket{\psi_0(t)}
\label{shrod_mod}
\end{equation}
being $\hat U \ket{\psi_0(t)}$ a source term (relevant when the potential is different from zero). By construction, we have the initial condition $\ket{\phi(0)}=0$. 

The decomposition \eqref{sum} can be very useful to simulate injection of particles into $\Omega\rq{}$ if we are able to analytically determine $\braket{x}{\psi_0}$ because we would not need to calculate it outside $\Omega\rq{}$. In fact, $\braket{x}{\phi}$ starts to become different from $0$ only when a non negligible part of $\braket{x}{\psi_0}$ interacts with the potential, i.e., when $\braket{x}{\psi_0}$ arrives inside $\Omega\rq{}$. By means of absorbing layers $\delta\Omega_{out}$ we can cancel out the part of $\braket{x}{\phi}$ that starts to flow out of $\Omega\rq{}$. Thus, the aim of the next section is to derive a unified analytical $\braket{x}{\psi_0}$ that works with both effective mass and tight binding Hamiltonians.

\subsubsection{Unified Gaussian free-particle evolution}\label{unified_section}

From literature, we only know the analytical $\psi_0(x,t)=\braket{x}{\psi_0(t)}$ solution of effective-mass Hamiltonian for a free particle in flat potentials, see Eq. \eqref{gaussian}. For the 1D atomistic tight binding Hamiltonian, we does not have an analytical solution for free particle Hamiltonian, however we can derive an approximate analytical solution that accurately works within many simulation cases of interest. 

We consider the initial state for $\braket{x}{\psi_0(t)}$ given by \eqref{gaussian_tb} and since eq.~\eqref{h_tbx}, assuming $\hat U=0$ and omitting for simplicity the subscript 0 of $\psi_0$ and the dependence on $t$, we have:
\begin{equation}
\brainket {j} {\hat H_0} {\psi}=  u (\psi_{j-1} -2 \psi_{j}+ \psi_{j+1})  + (\rho+2u)\psi_{j}.
\label{h_0x}
\end{equation}
where, from section \ref{section_hamiltonian}, $\psi_j=\braket{j}{\psi}=\psi(x_j)$ and $\psi_{j\pm 1}=\braket{j\pm 1}{\psi}=\psi(x_j\pm \Delta x)$ with $x_j$ the $j$--atom position and $\Delta x$ the distance between atoms.
Thus, using the Taylor series  $\psi_{j+1}=\sum_{r=0}^{\infty}\frac{1}{r!}\frac{\partial^r}{\partial x^r}\psi_j\Delta x^r$ (where $\frac{\partial^r}{\partial x^r}\psi_j=\frac{\partial^r}{\partial x^r}\psi(x)|_{x=x_j}$) and similarly for $\psi_{j-1}$ and substituting into \eqref{h_0x} we have
\begin{equation}
%\fracd{\psi_{j+1}-2\psi_j+\psi_{j-1}}{\Delta x^2} =\partialdersec{\psi_j}{x}+\fracd{\Delta x^2}{12}\fracd{\partial^4\psi_j}{\partial x^4}+O(\Delta x^4).
{\psi_{j+1}-2\psi_j+\psi_{j-1}} =2\sum_{r=1}^\infty\fracd{1}{(2r)!}\fracd{\partial^{2r}\psi_j}{\partial x^{2r}}\Delta x^{2r}.
\label{discretization_x_1}
\end{equation}
Now, looking at \eqref{gaussian}, it is a product of two exponentials, however the first exponential contains the stronger spatial variation of $\psi_0$, so we can neglect the spatial derivative of the second exponential and we get 
\begin{equation}
\fracd{\partial^{2r}\psi_G(x)}{\partial x^{2r}}\approx(ik_x)^{2(r-1)}\partialdersec{\psi_G(x)}{x},
\label{discretization_x_2}
\end{equation}
and, since we have assumed that the initial state for $\braket{x}{\psi_0(t)}$ is given by \eqref{gaussian_tb}, we can substitute into \eqref{discretization_x_1} and we have
\begin{equation}
%\fracd{\psi_{0_{j+1}}-2\psi_{0_j}+\psi_{0_{j-1}}}{\Delta x^2} \approx\left(1-\fracd{k_x^2}{12}\Delta x^2\right)\partialdersec{\psi_{0_j}}{x}.
\psi_{G_{j+1}}-2\psi_{G_j}+\psi_{G_{j-1}}\approx-\fracd{2}{k_x^2}\partialdersec{\psi_{G_j}}{x}\sum_{r=1}^\infty\fracd{(ik_x\Delta x)^{2r}}{(2r)!},
\label{discretization_x_3}
\end{equation}
and using the Taylor series of cosine in the r.h.s of \eqref{discretization_x_3} we have 
\begin{equation}
\psi_{G_{j+1}}-2\psi_{G_j}+\psi_{G_{j-1}} \approx2\fracd{1-\cos(k_x\Delta x)}{k_x^2}\partialdersec{\psi_{G_j}}{x}.
\label{discretization_x_final}
\end{equation}
Finally, substituting \eqref{discretization_x_final} into \eqref{h_0x} and defining the new time 
\begin{equation}
t\rq{}(t)=-\fracd{4um^*(1-\cos(k_x\Delta x))}{\hbar^2 k_x^2}t,
\label{time_free}
\end{equation}
it is simple to prove that the function 
\begin{equation}
\psi_{an}(x,t)=e^{-it(\rho+2u)/\hbar}\psi_G(x,t\rq{}(t)),
\label{analytical}
\end{equation}
under the above approximations, satisfies the tight binding version of equation \eqref{shrod} for $\hat U=0$. 

We can easily prove that the analytical solution \eqref{analytical} is not only a unified solution for both effective mass and tight binding free particle equation \eqref{shrod} with initial condition $\psi_G(x,0)$, but it is also valid for discretized version of the effective mass Hamiltonian commonly used in numerical simulations  
\begin{equation}
\brainket {j} {\hat H_0} {\psi}=-\fracd{\hbar^2}{2m^*}\fracd{\psi_{j+1}-2\psi_j+\psi_{j-1}}{\Delta x^2},
\label{discretization_x}
\end{equation}
in fact, \eqref{discretization_x} corresponds to \eqref{h_0x} when we consider $\rho=-2u$ and $u=-\hbar^2/(2m^*\Delta x^2)$.
Moreover, if we further consider small $k_x$ we have $t\rq{}\approx t$ and $E_{tb}(k)=\hbar^2 k^2/(2m^*)=E_{em}(k)$, meaning that only close to the bottom of the conduction band, the tight-binding model coincides with the effective-mass theory. 

\begin{figure}
\centerline{
\includegraphics[width=\columnwidth]{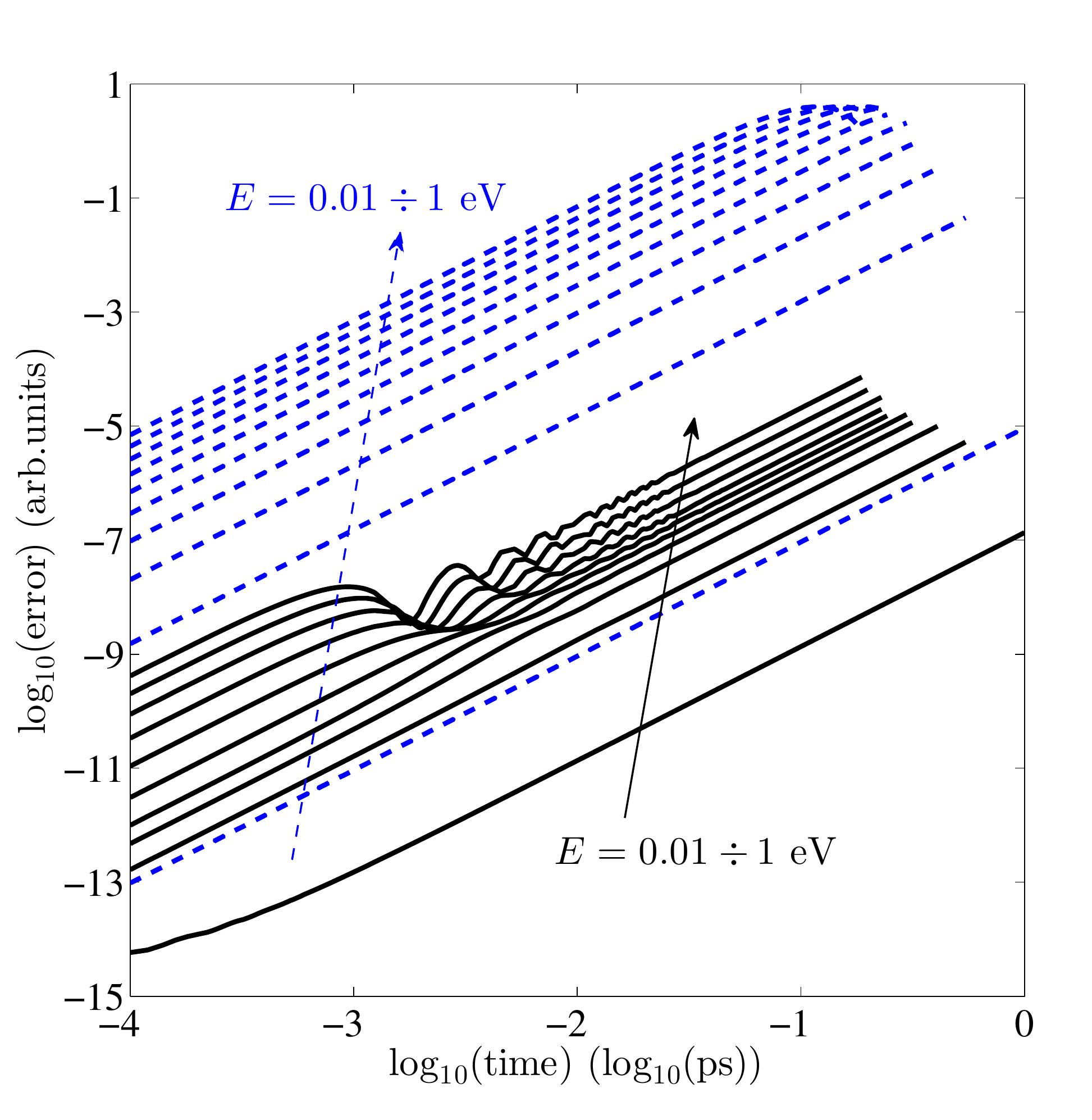}}
\caption{\label{err_compare_free_g} (Color online) Simulation of a Gaussian wave packet of a free particle with $E=\hbar^2 k_x^2/(2m^*)=0.01\div1$~eV equally spaced into 10 values, $\sigma_0=25/\sqrt{2}$~nm and initially centered in $x_0=-70$~nm and $m^*=0.2m_0$ where $m_0$ is the free electron mass.  The simulation parameters are $\Delta t=0.01$~fs, $\Delta x=0.2$~nm, the full simulation space ranges from $-800$ to $800$~nm, the simulation is stopped when $\int_{-800~\text{nm}}^{50~\text{nm}}|\psi_0|^2dx<10^{-10}$. The dashed blue line is $\log(\norm{\psi_{G_{num}}(t)-\psi_G(t)}^2)$, and the solid black line is $\log(\norm{\psi_{G_{num}}(t)-\psi_{an}(t)}^2)$ where $\psi_{G_{num}}$ is the numerical solution for the gaussian free particle, $\psi_G$ is the analytical solution \eqref{gaussian} and $\psi_{an}$ the analytical solution given by \eqref{analytical}.}
\end{figure}

In figure \ref{err_compare_free_g} the solution \eqref{analytical} is compared with the numerical solution for $\rho=-2u$ and $u=-\hbar^2/(2m^*\Delta x^2)$ at different values of $k_x$ obtained by inverting the relation $E(k_x)=\hbar^2 k_x^2/(2m^*)$. It can be seen that it leads to excellent agreement between the numerical solution and the analytical one even for high energies.

%%%%%%%%%%%%%%%%%%%%%%%%%
%%%%%%%%%%%%%%%%%%%%%%%%%%%%%%%%%%%%%%%%%%%%%%%%%%%%%%%%%%%%%%%%%%%%%%%%%%%
%%%%%%%%%%%%%%%%%%%%%%%%%
%%%%%%%%%%%%%%%%%%%%%%%%%

\subsection{Absorption}\label{section_dumping}

%%%%%%%%%%%%%%%%%%%%%%%%%
%%%%%%%%%%%%%%%%%%%%%%%%%
%%%%%%%%%%%%%%%%%%%%%%%%%

The injection algorithm avoids the simulation of the quantum state outside $\Omega\rq{}$ before and in a short time after the interaction occurring within $\Omega\rq{}$. However, after a time large enough (but much smaller than typical simulation time), the quantum state $\braket{x}{\phi}$ defined in section \ref{section_state_split} starts to spread out $\Omega\rq{}$. Thus, in order to avoid the simulation out of $\Omega\rq{}$, we are interested in a function $\Psi(x,t)$ that would be equal to the solution $\psi(x,t)$ of Eq. (\ref{shrod}) in $\Omega\rq{}$, i.e., $\Psi(x,t)=\psi(x,t)$ at $x \in \Omega'$, but that it could vanish outside, i.e., $\Psi(x,t) \approx 0 $ for $x\not\in \Omega\rq{}$. In order to achieve this goal, we discuss a modified version of a well known absorbing algorithm \cite{Kosloff}. 

\begin{figure}
\centerline{
\includegraphics[width=.9\columnwidth]{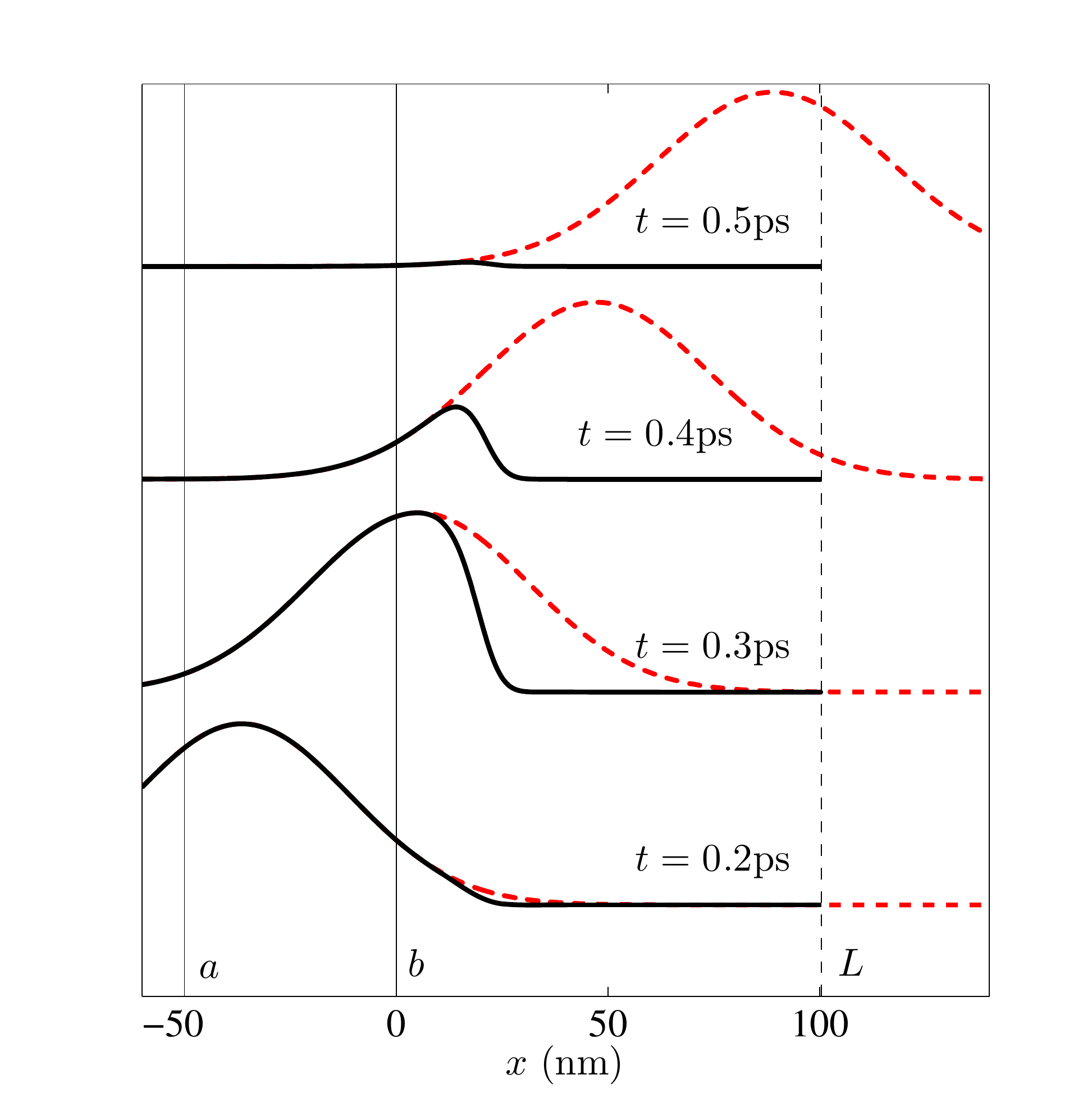}}
\caption{\label{absorb1} (Color online) Simulation of gaussian wave packet with energy $E=0.1$~eV and the other parameters as in figure \ref{err_compare_free_g}. The solid black line  is the absorbed wave packet with $L=100$~nm and the dashed red line is the wave packet given by \eqref{gaussian}.}
\end{figure}

Let us consider $\Omega\rq{}$ defined by $a\leq x\leq b$ with $a<b$ and $a,b\in\mathbb{R}$. For $x\geq b$ and $x\leq a$ the potential $U$ is assumed uniform.  With no loss of generality, we take $b=0$ and we discuss the boundary condition for $x\geq b$ only. We define the function $f$ as
\begin{equation}
f(x)=\left\{
\begin{array}
[c]{ll}
g(x) & \text{ \ for }x>0\\
1 & \text{ \ for }x\leq0
\end{array}
\right.  
\label{f}
\end{equation}
with $g(x)$ a real positive smooth function smaller than $1$. The goal is to define a recursive algorithm to make $\psi$ vanishing in a region $0<x\leq L$ (the absorbing layer $\delta\Omega_{out}$) for some $L>0$ without perturbing the part of the wave function belonging to $\Omega\rq{}$. Using the central difference scheme to integrate \eqref{shrod} the first iteration reads
\begin{equation}
\psi^2=\psi^0+ \frac{\Delta t}{i\hbar}\hat H \psi^1
\end{equation}
where $\hat H$ is the Hamiltonian $\hat H=-\frac{\hbar^2}{2m}{\frac{\partial^2}{\partial x^2}}+U$ and $\psi^j=\psi(x,t_j)$. Let  $\Psi^0=f\psi^0$ and $\Psi^1=f\psi^1$ and we modify the first iteration as
\begin{equation}
\Psi^2=\Psi^0+ \frac{\Delta t}{i\hbar}\hat H \Psi^1=f\psi^0+ \frac{\Delta t}{i\hbar}\hat H f \psi^1
\end{equation}
Now, we further assume that $f(x)$ is sufficiently smooth to commute with $\hat H$, $[f,\hat H] \approx 0$,  obtaining
\begin{equation}
\Psi^2=f\left(\psi^0+ \frac{\Delta t}{i\hbar}\hat H \psi^1\right)
\end{equation}
Iterating the scheme and using (\ref{f}) we have
\begin{equation}
\Psi^n=f^{n-1}\psi^n=\left\{
\begin{array}
[c]{ll}%
g(x)^{n-1}\psi(x,t_n) & \text{ \ for }x>0\\
\psi(x,t_n) & \text{ \ for }x\leq0
\end{array}
\right.
\label{schrod_tn}
\end{equation}
Unfortunately, the function $g(x)$ rigorously satisfying all the previous prescriptions does not exist. Indeed the unique real function that commutes
with $\hat H$ and satisfies all the analytical properties stated above is $g(x)=1$,
but it does not satisfy $g(x)<1$. However we can require a function that only approximately commutes with $\hat H$. This weaker condition can be reached requiring that both the first and second spatial derivatives of $g(x)$ are small enough compared to the spatial derivatives of $\psi$ where $\psi$ is not negligible.
Among many other possibilities, we can use a slightly decreasing polynomial of the form
\begin{equation}
g(x)=1-\left(\frac {x} {L} \right)^m\label{gpol}
\end{equation}
with $m\geq3$. The polynomial \eqref{gpol} for $0\leq x\ll L$ has very small derivatives, so that it approximately commutes with $\hat H$ as required. When $x$ approaches to $L$ the derivatives of $g(x)$ increase, however, as shown in figure~\ref{absorb1} the wave function is absorbed much before $x=L$. Finally, from figure~\ref{absorb1} it can be seen that the wave-function is not perturbed inside $\Omega\rq{}$, as required. 

From equation \eqref{gpol} follows that the boundary condition can be modulated by varying both $L$ and $m$. In order to automatically estimate $L$ we observe that for a Gaussian wave packet the characteristic length is the de Broglie length $\lambda=2\pi/k_x$. It is simple to see that if $L\gg\lambda$ then $f(x)$ approximately commutes with $\hat H$. One can use this argument to define $L$. Other criteria for fixing $L$ that satisfy a predetermined error are also possible. In any case, one expects that wave functions with high energies requires a smaller $L$ than that of low energy wave functions. It is worth noticing that for zero applied bias, if in $\Omega\rq{}$ the wave packet interacts with some potential barrier, the transmitted and reflected waves have momenta that are in general close to the initial one (a part from a sign) so, the length $L$ can be simply related to the initial de Broglie length. When a bias is applied, there is an asymmetry  between the right and left wave lengths of the wave packet that needs to be taken into account. 

\begin{figure}
\centerline{
\includegraphics[width=\columnwidth]{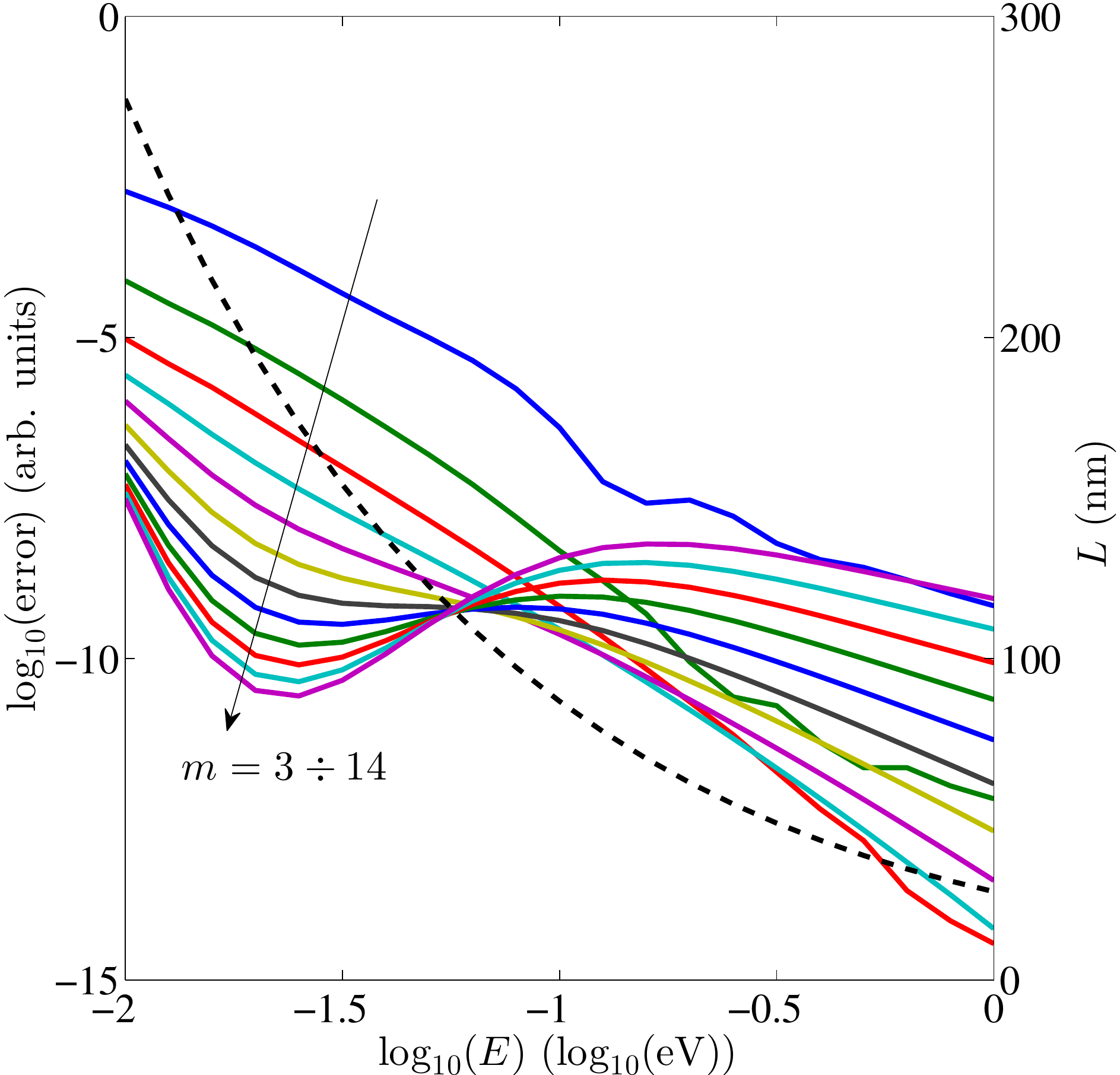}}
\caption{\label{err_compare_n} (Color online) Simulation of a Gaussian wave packet of a free particle with the same parameters of figure \ref{err_compare_free_g}. The simulation parameters are: the full simulation space ranges from $-800$ to $800$~nm, $a=-800$~nm, $b=50$~nm, the length $L$ taken as $10\lambda$ is the dashed black line referred to the right $y$--axis. The solid lines are $\log_{10}(\varepsilon_{abs})$.}
\end{figure}

We have carried out simulations to determines the impact of $m$ on the error of the absorbing argument. In Fig. \ref{err_compare_n} we evaluated the error function $\varepsilon_{abs}=\max_{t\in \mathbb{R}_+}[\int^b_a|\psi_{G_{num}}(t)-\Psi_{G_{abs}}(t)|^2dx)]$ where $\psi_{G_{num}}$ is the numerical solution of the Schr\"{o}dinger equation calculated in the full spatial domain, i.e. the domain $\Omega$ in figure \ref{domain}(a), and $\Psi_{G_{abs}}$ is the solution obtained for the wave function absorbed with $L$ depending on the initial energy of the wave packet. Thus, looking at Fig. \ref{err_compare_n}, the best compromise for $m$, i.e., best behavior for high and low energies, is $5\leq m\leq 7$. 

We finally mention that this absorbing algorithm is equivalent to introduce in the equation \eqref{shrod} a negative imaginary potential non-vanishing outside $\Omega\rq{}$ as discussed in appendix \ref{appendix1}.

%%%%%%%%%%%%%%%%%%%%%%%%%
%%%%%%%%%%%%%%%%%%%%%%%%%
%%%%%%%%%%%%%%%%%%%%%%%%%

\subsection{Remapping}\label{section_remapping}

%%%%%%%%%%%%%%%%%%%%%%%%%
%%%%%%%%%%%%%%%%%%%%%%%%%
%%%%%%%%%%%%%%%%%%%%%%%%%

As mentioned above in the results of Fig. \ref{err_compare_n}, wave functions with small energies requires large absorbing layers $\delta\Omega_{out}$.  For such small energies one can expect a reduction of the error, without increasing the number of grid points, by a proper remapping of the absorbing layers.  Let us consider the same simulation domain of the previous section with $\Omega\rq{}$ defined by $a\leq x\leq 0$. For $x\geq 0$ we define the variable change 
\begin{equation}
z=c(x)=K\arctan{\fracd{x}{K}}.
\label{variable_change}
\end{equation}
This maps $x$ into $z$ with a consequent contraction of the spatial domain, in fact for $x\in[0,+\infty]$ we have $z\in[0,K\pi/2]$. Moreover, deriving \eqref{variable_change} we have $dz/dx=K^2/(K^2+x^2)$ that in $x=0$ is equal to $1$ meaning that the contraction map is smooth in the whole spatial domain. We can unambiguously define the inverse map for $z\in[0,K\pi/2]$ as
\begin{equation}
x=c^{-1}(z)=K\tan{\fracd{z}{K}}.
\label{inverse_vc}
\end{equation}
Using \eqref{variable_change} and \eqref{inverse_vc} we can rewrite the Schr\"{o}dinger equation \eqref{shrod} for $x\in[0,+\infty]$ with wave-function $\psi(c^{-1}(z),t)$ using the transformed Hamiltonian 
\begin{align}
\hat H_z=-\fracd{\hbar^2}{2m^*}\left[c\rq(c^{-1}(z))^2\partialdersec{}{z}+c\rq{}\rq(c^{-1}(z))\partialder{}{z}\right]\nonumber\\
+U(c^{-1}(z),t),
\label{H_remapped}
\end{align}
where $c\rq(x)=dc(x)/dx=K^2/(K^2+x^2)$ and $c\rq{}\rq(x)=d^2c(x)/dx^2=-2K^2x/(K^2+x^2)^2$. Using \eqref{inverse_vc} these expressions give $c\rq(c^{-1}(z))^2=\cos^4(z/K)$ and $c\rq{}\rq(c^{-1}(z))=-2/K\sin (z/K)\cos^3(z/K)$.

\begin{figure}
\centerline{
\includegraphics[width=\columnwidth]{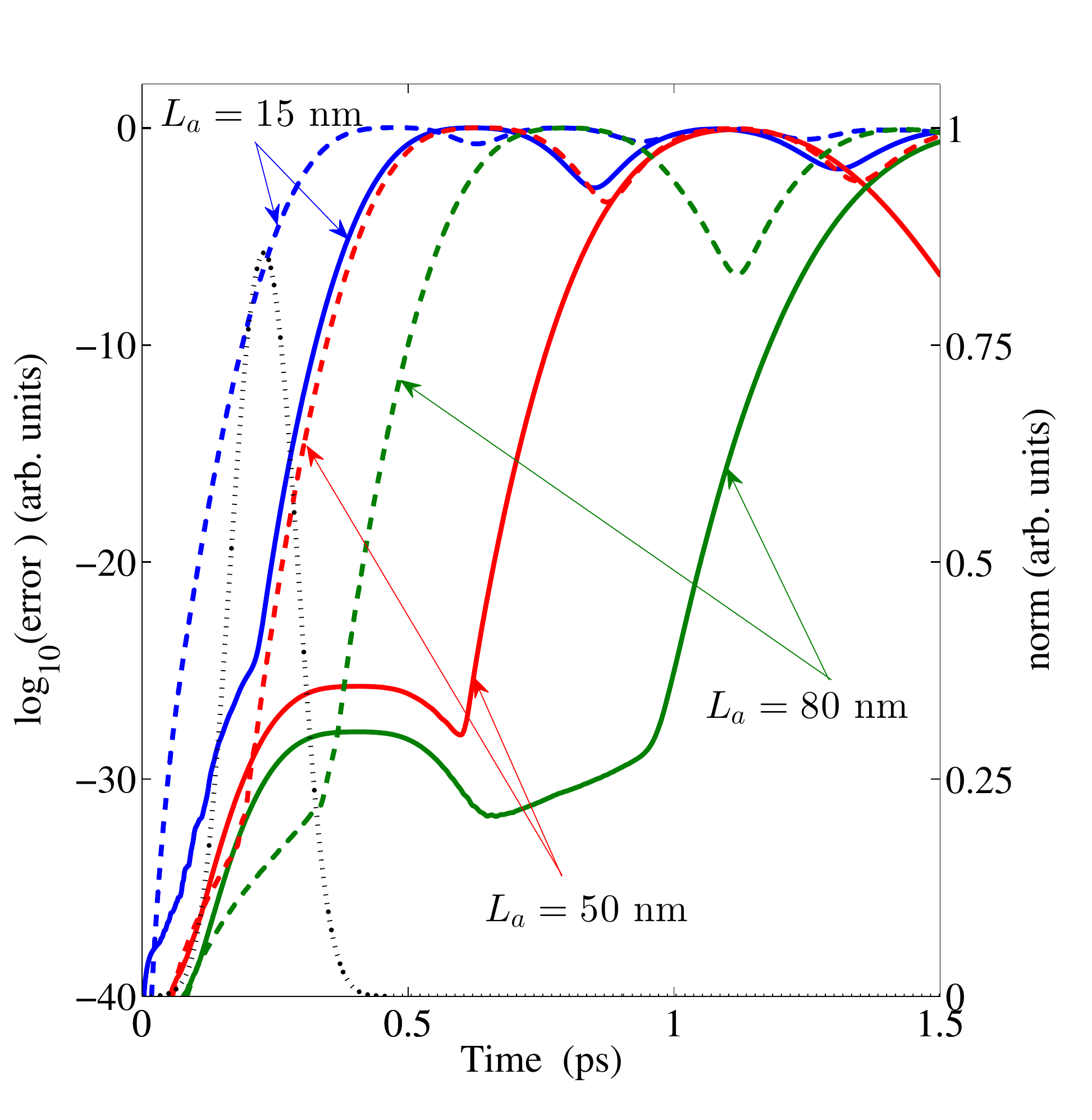}}
\caption{\label{err_compare_remap} (Color online) Simulation of a Gaussian wave packet of a free particle with the same parameters of figure \ref{err_compare_free_g}. The simulation parameters are: $E=0.1$~eV, the full simulation space ranges from $-800$ to $800$~nm, $a=-800$~nm, $b=50$~nm. The dotted black line referred to the right $y$--axis is the norm $\int_{-800~\text{nm}}^{50~\text{nm}}|\psi_G|^2dx$. The solid lines are $\log_{10}(\varepsilon_{rem})$ and the dashed lines are $\log_{10}(\varepsilon_{cut})$.}
\end{figure}

Let us define $L_a$ the width of the spatial domain outside $\Omega\rq{}$, so the spatial domain is divided into $a-L_a\leq z< a$ (left augmented boundary), $a\leq x\leq b$ ($\Omega\rq{}$) and $b< z\leq b+L_a$ (right augmented boundary). Thus, the remapping parameter $K$ of \eqref{variable_change} is assumed to be  equal for both sides because the augmented boundaries are the same. It is worth noticing that, for a correct implementation we take $K=2L_a/\pi$ implying $c^{-1}(a-L_a)=-\infty$ and $c^{-1}(b+L_a)=\infty$. In order to discuss the numerical results when we implement the remapping algorithm, we assume $a=-\infty$ so no injection is needed. We define $\varepsilon_{rem}=\int_a^b|\psi_{full}-\psi_{rem}|^2dx$ where $\psi_{full}$ is the numerical solution of the free particle in an infinite (large enough) domain and $\psi_{rem}$ the numerical solution using the remapping on the right side of the domain. Moreover we define $\varepsilon_{cut}=\int_a^b|\psi_{full}-\psi_{cut}|^2dx$ where $\psi_{cut}$ is the numerical solution in the domain $-\infty< x\leq b+L_a$.
From figure \ref{err_compare_remap}, it is worth noticing that, when we implement the remapping, after a certain delay the error $\varepsilon_{rem}$ starts growing. This is due to the fact that when we numerically implement the Hamiltonian \eqref{H_remapped}, the differential part corresponds to a discretized second derivative in $x$ with a $\Delta x$ growing as $\Delta x=K \tan(\Delta z/K)$ (we used the \eqref{inverse_vc}). Since the considerations done in section \ref{unified_section}, it results in a slowdown of the wave function. However, when $K \tan(\Delta z/K)$ grows too much, the wave function is reflected and it returns back to $\Omega\rq{}$. So basically we have the same behavior of a wave function simulated in the domain $-\infty< x\leq b+L_a$ (see the error $\varepsilon_{cut}$ in figure \ref{err_compare_remap}) with the unique difference that the reflection is retarded. 

%%%%%%%%%%%%%%%%%%%%%%%%%
%%%%%%%%%%%%%%%%%%%%%%%%%
%%%%%%%%%%%%%%%%%%%%%%%%%

\section{Practical implementation}\label{section_implementation}

%%%%%%%%%%%%%%%%%%%%%%%%%
%%%%%%%%%%%%%%%%%%%%%%%%%
%%%%%%%%%%%%%%%%%%%%%%%%%

Starting from the considerations in the previous sections, we can devise a new algorithm overcoming all the drawbacks reported above. The idea is to combine all the previous algorithms in such a way they compensate their drawbacks. The simultaneous employment of those algorithms is aimed to simulate the Schr\"{o}dinger equation using as spatial support $\Omega\rq{}$ plus small absorbing layers $\delta\Omega_{out}\rq{}$. In table \ref{algorithm} a pseudo-code for simultaneous implementation is reported. We discuss here the main idea and some implementation details to improve the accuracy of the simulation.
 
\begin{table}[t]
\boxed{
\begin{minipage}[c][10.2cm][t]{0.96\columnwidth}
\%\textit{Main definitions}\\
\raggedright
\texttt{ib = a$\leq$x$\leq$b } \hspace{3.6cm}\%\textit{Points inside $\Omega\rq{}$}\\
\texttt{il = x<a,  ir = x>b }  \hspace{2.3cm}\%\textit{Points outside $\Omega\rq{}$}\\
\texttt{K = 2*La/pi}  \hspace{2.95cm}\%\textit{Remapping parameter}\\
\texttt{L = 10*(2*pi/kx)}  \hspace{2.15cm}\%\textit{Absorption parameter}\\
\texttt{tc = 2*(1-cos(kx*dx))/(kx*dx)\^{}2}  \hspace{0.1cm}\%\textit{Time correction}\\
\texttt{}  \hspace{6.95cm}\textit{constant}\\
\texttt{zl = a+K*tan((x(il)-a)/K)}  \hspace{1.5cm}\%\textit{Variable change}\\
\texttt{zr = b+K*tan((x(ir)-b)/K)}  \hspace{1.5cm}\%\textit{Variable change}\\
\texttt{g(il) = 1-(a-zl)\^{}n*L\^{}-n} \hspace{1.95cm}\% \textit{$f(x)$ for $x<a$}\\
\texttt{g(ir) = 1-(zr-b)\^{}n*L\^{}-n} \hspace{1.95cm}\% \textit{$f(x)$ for $x>b$}\\
\centering
\%\textit{Cycle over time steps}\\
\raggedright
\texttt{do j = 1..N}\\ 
\centering
\%\textit{Solve for $\phi$}\\
\raggedright
\hspace{0.5cm}\texttt{phi2(il) = g(il).*(phi0(il)+Hzl*phi1(il)+}\\
\raggedleft
\texttt{U(zl,t\_j).*(phi1(il)+psi1\_0(il)))}\\
\raggedright
\hspace{0.5cm}\texttt{phi2(ib) = phi0(ib)+H*phi1(ib)+}\\
\raggedleft
\texttt{U(ib,t\_j).*(phi1(ib)+psi1\_0(ib))}\\
\raggedright
\hspace{0.5cm}\texttt{phi2(ir) = g(ir).*(phi0(ir)+Hzr*phi1(ir)+}\\
\raggedleft
\texttt{U(zr,t\_j).*(phi1(ir)+psi1\_0(ir)))}\\
\raggedright
\hspace{0.5cm}\texttt{phi0 = phi1,    phi1 = phi2}\\
\centering
\%\textit{Solve for $\psi_0$}\\
\raggedright
\hspace{0.5cm}\texttt{psi2\_0(il) = psi0\_an(zl,t\_j*tc)}\\
\hspace{0.5cm}\texttt{psi2\_0(ib) = psi0\_0(ib)+H*psi1\_0(ib)}\\
\hspace{0.5cm}\texttt{psi2\_0(ir) = g(ir).*(psi0\_0(ir)+Hzr*psi1\_0(ir)}\\
\hspace{0.5cm}\texttt{psi0\_0 = psi1\_0,    psi1\_0 = psi2\_0}\\
\texttt{}\\
\texttt{end do}
\end{minipage} 
}
\caption{\label{algorithm}Pseudo-code for simultaneous implementation of injection, absorption and remapping algorithms. Product \texttt{.*} denotes element-by-element multiplication between vectors.}
\end{table}

Let $L$ and $L_a$ as in section \ref{section_dumping} and section \ref{section_remapping} respectively. We further define an effective augmented length $L_{eff}\leq L_a$ such that $L_{eff}=K\arctan L/(2K)$ and it is the width of the absorbing layer $\delta\Omega_{out}\rq{}$. So, in this picture, the spatial domain is now divided into $a-L_{eff}\leq z< a$ ($\delta\Omega_{out}\rq{}$ on the left), $a\leq x\leq b$ ($\Omega\rq{}$) and $b< z\leq b+L_{eff}$ ($\delta\Omega_{out}\rq{}$ on the right). Thus, $L_a$ is, in this case, the maximum augmented boundary when $L=\infty$. The remapping parameter $K$ is yet evaluated through $L_a$ as in section \ref{section_remapping}. The absorbing parameter $L$ has to be chosen enough larger than the de Broglie length to guarantee $L\gg\lambda$. Here we take $L=10\lambda$. It is worth noticing that the function $g(x)$ has been consistently evaluated with the remapping to guarantee the efficiency of the absorption algorithm, i.e., the equation \eqref{gpol} must be remapped through \eqref{inverse_vc} (see table \ref{algorithm}). 

In this picture, we assume for simplicity that the particle is injected from the left reservoir (injection from the right follows straightforwardly). We use the injection algorithm discussed in section \ref{section_Injection}. In table \ref{algorithm} the evaluation of $\psi_0$ is analytical only in part. In fact, assuming that the packet comes from the left reservoir, it is  analytically evaluated only there following Eq.~\eqref{gaussian} and remapped through \eqref{inverse_vc}. Even if in the rest of the domain $\psi_0$ is numerically evaluated, this does not increase the numerical burden because the number of floating point operations are practically the same. On the contrary this choice permits to consistently implement the absorption algorithm also for $\psi_0$ that from numerical tests results in a more accurate solution. 

\begin{figure*}
\centerline{
\includegraphics[width=2.05\columnwidth]{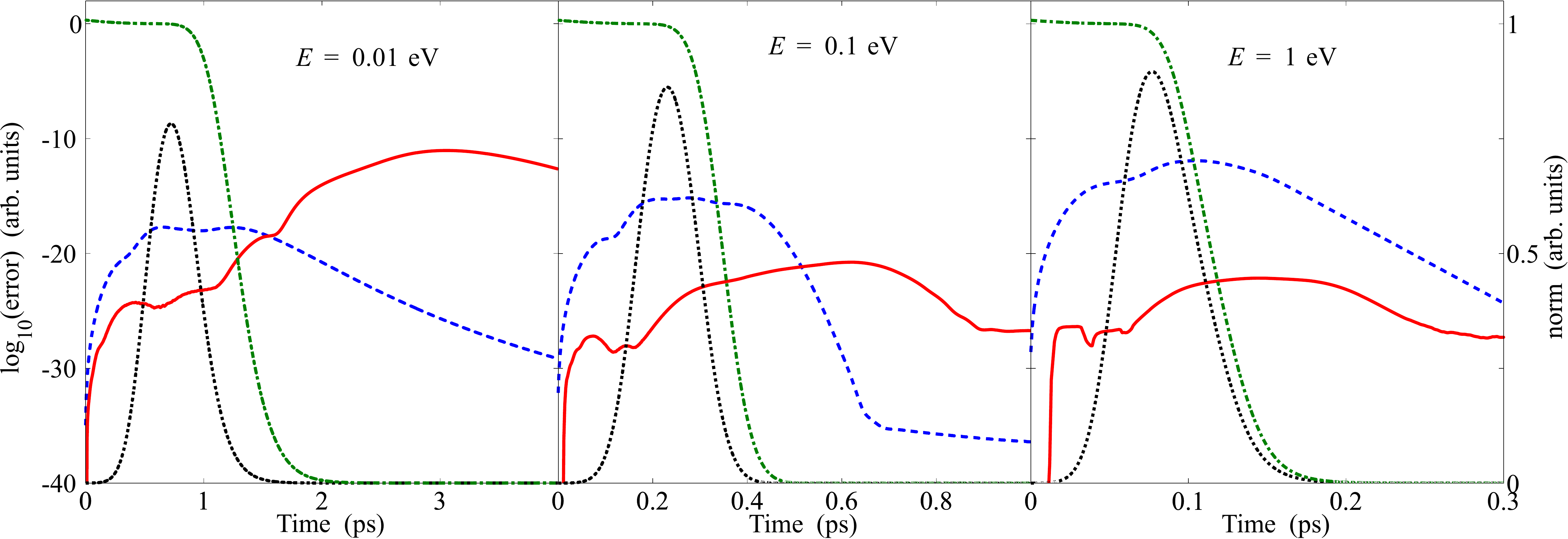}}
\caption{\label{err} (Color online) Simulation of a gaussian wave packet with the same parameters of figure \ref{err_compare_free_g}. The simulation is carried out for the three energies $E=0.01$, $0.1$ and $1$~eV, the wave-packet interacts with a potential barrier centered at $27.5$~nm with width of $5$~nm and heights of $0.015$, $0.0825$ and $0.93$~eV respectively. The different barrier heights guarantee that the transmitted and reflected waves after interaction have the same norm for each energy guaranteeing that the absorbing and remapping algorithm equally work on both sides of $\Omega\rq{}$. The full simulation space ranges from $-800$ to $800$~nm, $\Omega\rq{}$ from $a=0$~nm to $b=50$~nm, the absorbing layers have $L_a=20$~nm, $L=274$, $86.8$ and $27.4$~nm (corresponding to double of the width of $\delta\Omega_{out}$) and $L_{eff}=18.9$, $16.4$, $10.5$~nm (corresponding to the width of $\delta\Omega_{out}\rq{}$) respectively depending on the initial wave packet energy and finally the absorption exponent is $m=5$. The dashed blue line is $\log(\varepsilon_{inj})$ and the solid red line is $\log(\varepsilon_{ar})$ and they are referred to the left $y-$axis. The dotted black line is the evolution of the norm $\int_a^b|\psi_{num}|^2dx$ and the dashed-dotted green line is the evolution of the norm $\int_{a-L_{a}}^{a}c^{-1}(z)|\Psi_{new}|^2dz+\int_{a}^{b}|\Psi_{new}|^2dx+\int_{b}^{b+L_{a}}c^{-1}(z)|\Psi_{new}|^2dz$ and it is referred to the right $y-$axis.}
\end{figure*}

The remapping and absorbing algorithms are used simultaneously into the absorbing layers. The first has as practical effect the slowdown of the wavefunction inside the absorbing layers while the absorbing algorithm cuts down the wave function avoiding the reflection discussed in section \ref{section_dumping}. This allows us to use very small absorbing layers to simulate the wave function. In order to give a practical example, in figure \ref{err} we reported the simulation of a particle interacting with a potential barrier for three different energies $E=0.01$, $0.1$ and $1$~eV. The direct implementation of the absorbing  algorithm would require the absorbing layers $\delta\Omega_{out}$ large at least $L/2=137$, $43.4$ and $13.7$~nm respectively (see section \ref{section_dumping}). On the contrary, using the remapping algorithm, we can take $L_a=20$~nm, and as consequence the with of $\delta\Omega_{out}\rq{}$ is $L_{eff}=18.9$, $16.4$, $10.5$~nm resulting in a much smaller absorbing layers specially for small energies.

When we solve for $\phi$ and $\psi_0$ (where the solution is not analytical) in table \ref{algorithm} we used a central difference scheme for the time derivative that is an explicit method, stable for Schr\"{o}dinger Equation, widely used and exhaustively discussed in \cite{Askar78}. However the choice is not mandatory and the entire algorithm can be simply modified according to other finite difference schemes for the time derivative. The Hamiltonians $H$ and $H_z$ can be evaluated using normal finite difference schemes. In table \ref{algorithm}, $H$, $H_z$ and $U$ are implicitly multiplied by $2\Delta t/(i\hbar)$. 

Finally in Fig. \ref{err} the error of the method is reported. We simulated a Gaussian wave packet interacting with a potential barrier. We have defined three different solutions to estimate different sources of errors. The first is $\psi_{num}$: the full numerical solution obtained using a large enough spatial domain to avoid boundary effects.  The second is $\psi_{inj}$: the solution with large spatial domain including the injection algorithm implemented as in table \ref{algorithm}. The last is $\Psi_{new}$, solution in the small domain including all the algorithms implemented as in table \ref{algorithm}. Using these different solutions, we define the errors $\varepsilon_{inj}=\int^b_a|(\psi_{num}-\psi_{inj})|^2dx$ due to the injection algorithm only and $\varepsilon_{ar}=\int_a^b|(\psi_{inj}-\Psi_{new})|^2dx$ due to the absorbing and remapping algorithm simultaneously used. As it can be seen from the caption of Fig. \ref{err} the small domain $\delta\Omega_{out}\rq{}\cup\Omega\rq{}\cup\delta\Omega_{out}\rq{}$ (of width $50+2L_{eff}$~nm) is much smaller than the full domain $\Omega$ ($1600$~nm), about two orders of magnitude, and than the domain $\delta\Omega_{out}\cup\delta\Omega_{inj}\cup\Omega\rq{}\cup\delta\Omega_{out}$ (of width $50+200+2L/2$), about one order of magnitude, resulting in an extreme reduction of the computational burden. The error $\varepsilon_{inj}$ is always very small (negligible) for any energy. Coupling the absorbing algorithm to the remapping we contract the entire $x$ domain inside the augmented boundary and consequently the absorbing algorithm can properly work as $\varepsilon_{ar}$ proves. Finally we observe that the total error of our algorithm follows $\varepsilon_{tot}\leq\varepsilon_{inj}+\varepsilon_{ar}$ that in $\log$ scale of figure \ref{err} means that it is the maximum between them.

%%%%%%%%%%%%%%%%%%%%%%%%%
%%%%%%%%%%%%%%%%%%%%%%%%%
%%%%%%%%%%%%%%%%%%%%%%%%%

\section{Conclusions}

%%%%%%%%%%%%%%%%%%%%%%%%%%
%%%%%%%%%%%%%%%%%%%%%%%%%%
%%%%%%%%%%%%%%%%%%%%%%%%%%

In the literature, most of the efforts to deal with quantum dynamics are still based on the use of Hamiltonian or momentum eigenstates. However, one of such states cannot describe an electron localized at the right or left of the barrier because they extend everywhere at any time \cite{footnote3}. In other words, Hamiltonian or momentum eigenstates (or scattering states) do not belong to the Hilbert space because they cannot be normalized to unity. Additionally, any attempt to include many-body physics in single-particle solutions (for example with the use of conditional Bohmian wavefunctions) do also require explicit time-dependent equations. However, the practical solution of such time-dependent equations faces with very important computational problems. One of the reasons of the poor development of explicit time-dependent quantum model is the difficulty for developing accurate and fast algorithms for quantum time-dependent equations. 

In this paper, we have presented three mathematical algorithms (analytical injection, absorption and remapping) that allow accurate and fast simulations for the time-dependent one-dimension of Schr\"{o}dinger equation with a spectacular reduction of the simulation box. We discuss the advantages and disadvantages of the three algorithms presented in this work:

\begin {itemize}
\item The analytical injection is extremely useful to simulate injection of particles, we only need to calculate $\psi_0$ inside $\Omega\rq{}$. Besides, there is no limitation of the initial position of the wave packet and, as we prove in the section \ref{unified_section}, this injection model can be used for the continuous as well as for the tight binding version of the Schr\"{o}dinger Equation.
\item By applying absorption boundary to the simulation box, we eliminate the spurious reflection by cutting down the wave function inside the augmented boundary. Moreover, we can shorten the simulation box without losing the accuracy of simulation. But the wave function with low energy requires larger simulation box than that of high energy wave function, which is shown in Fig.~\ref{err_compare_n}.
\item The remapping algorithm used alone has the unique effect to slowdown the wave function in the augmented boundary, but it does not avoid the reflection. However, when coupled with the absorbing boundary algorithm of section \ref{section_dumping} it compensates the drawback of the absorbing algorithm, i.e., it allows to greatly reduce the absorbing layer width even for low energies. 
\end{itemize}

Finally, we created a new model by simultaneously combining all three previous algorithms to simulate quantum transport, which worked well for both low and high energies wave functions. The spatial domain can be reduced (50+2$L_{eff}$ nm with, for example, $L_{eff}=10.5\; nm$) to less than 5 per cent of the original domain (1600 nm), while introducing an almost negligible error. Since the needed memory and the number of operations is proportional to the grid points, a similar percentage of reduction of memory and CPU time can be expected. The generalization to 2D (even 3D) systems can be done straightforwardly with the only requirement that the free evolution of the initial wave packet is well-known. For example, a just analytical solution of a Gaussian wave packet solution of the Dirac equation is needed to provide tight-binding simulations of 2D graphene \cite{graphene1,graphene2} with a similar strategy as done in section \ref{unified_section}. 

All the results developed here for quantum transport can also be applied to many other fields that requires a time-dependent solution of the Schr\"{o}dinger equation and whose initial states are analytically defined far from the interaction region \cite{Robinett_2011}.  Additionally, the injection and absorbing algorithms can also be applied only in one side of the system for those computations that require an explicit spatial knowledge of the transmitted wave packet far from the interaction region \cite{krammer_2010}. 

\begin{acknowledgments}

The authors acknowledge support from the \lq\lq{}Ministerio de Ciencia e Innovaci\'{o}n\rq\rq{} through the Spanish Project TEC2012-31330 and by the Grant agreement no: 604391 of the Flagship initiative  \lq\lq{}Graphene-Based Revolutions in ICT and Beyond\rq\rq{}. Z. Zhan acknowledges financial support from the China Scholarship Council (CSC).

\end{acknowledgments}

%Create the reference section using BibTeX:
%\bibliography{schrod_bound}

\appendix
\section{Negative imaginary potential}\label{appendix1}

In order to prove that the absorbing layer algorithm is equivalent to a negative imaginary potential in the absorbing layer we consider the one-dimensional time-dependent Schr\"{o}dinger equation \eqref{shrod} and the space artificially divided into two reservoirs (left and right) and $\Omega\rq{}$ as in section \ref{section_dumping}. Then, $\Omega\rq{}$ is defined by $a\leq x\leq b$ with $a<b$ and $a,b\in\mathbb{R}$. For $x\geq b$ and $x\leq a$ the potential $U$ is assumed uniform and for the sake of simplicity we take b=0 and discuss the boundary condition for $x\leq b$ only.

We consider the functions $f$ and $g$ defined in \eqref{f} and at the time $t=0$ we define $\Psi(x,0)=f(x)\psi(x,0)$ with $\psi(x,0)$ the initial state of \eqref{shrod}. As discussed in section \ref{section_dumping}, at each time step the wave function $\Psi(x,t)$ can be written as $\Psi (x,t)=f(x) \psi(x,t)$ iff $f(x)$ is sufficiently smooth to commute with Hamiltonian $\hat H$. Then at a finial time $t$ using \eqref{schrod_tn} we have
\begin{equation}
\Psi(x,t)=f(x)^{t/\Delta t}\psi(x,t),
\label{schrod_tn1}
\end{equation} 
where, being $\Delta t$ the temporal step, from equation \eqref{schrod_tn} $t/\Delta t=n-1$. Rewriting the equation\eqref{schrod_tn1}, we obtain
\begin{eqnarray}
\Psi(x,t)=\exp\left[{-\frac{1}{\hbar}\left(-\frac{\hbar}{\Delta t}\log[f(x)]\right)t}\right]\times\nonumber\\
\times\psi(x,t)=\exp\left[-\frac{J(x)}{\hbar}t\right]\psi(x,t),
\label{schrod_tn2}
\end{eqnarray}
where we define $J(x)=-\frac{\hbar}{\Delta t}\log[f(x)]$. Thus, inverting \eqref{schrod_tn2} we get 
\begin{equation}
\psi(x,t)=e^{\frac{J(x)}{\hbar}t}\Psi(x,t)
\label{psi_Psi}
\end{equation}
and inserting \eqref{psi_Psi} into \eqref{shrod}, we have
 \begin{equation}
i\hbar\partialder{}{t}e^{\frac{J(x)}{\hbar}t}\Psi(x,t)=\hat H e^{\frac{J(x)}{\hbar}t}\Psi(x,t).
\label{schrod_xtn1}
 \end{equation}
Since the function $f$ approximately commutes with $\hat H$, we also have $[e^{\frac{J(x)}{\hbar}t},\hat H] \approx 0$. Using this commutation rule, developing the time derivative and dividing by $e^{\frac{J(x)}{\hbar}t}$ the \eqref{schrod_xtn1} can be rewritten as
\begin{equation}
i\hbar \partialder{\Psi(x,t)}{t}=[\hat H- iJ(x)]\Psi(x,t),
\label{schrod_xt'n3}
\end{equation}
proving that the absorbing layer algorithm is equivalent to a negative imaginary potential.    

\bibliography{schrod_bound}

\begin{thebibliography}{37}
\bibitem [{foo()}]{footnote3}%
  \BibitemOpen
  \href@noop {} {}\bibinfo {note} {A Hamiltonian eigenstate has a
  time-independent probability $|\psi_E(x,t)|^2=\rho(x)$. Then, the continuity
  equation $\partial \rho(x)|\partial t +\nabla J=0$ implies that either the
  current density is zero $J=0$ (such selection is not valid for quantum
  transport) or $J$ is uniform everywhere so that its spatial derivative is
  zero. The latter means that $\psi_E(x,t)$ is a complex function that extended
  everywhere in the Hilbert space at any time.}\BibitemShut {Stop}%
\end{thebibliography}

\end{document}